\documentclass[twocolumn]{pasj00}

\usepackage{natbib, aas_macros}
\usepackage{bm}

\begin{document}
\SetRunningHead{Kawashima et al.}{Limit-Cycle Activities in Swift
J1644+57}
\Received{2013/4/16}
\Accepted{2013/5/14}

\title{Recurrent Outbursts and Jet Ejections Expected in Swift J1644+57:\\
Limit-Cycle Activities in a Supermassive Black Hole}

\author{%
  Tomohisa \textsc{Kawashima}\altaffilmark{1},
  Ken \textsc{Ohsuga}\altaffilmark{2,3},
  Ryuichi \textsc{Usui}\altaffilmark{4},
  Nobuyuki \textsc{Kawai}\altaffilmark{4},
  Hitoshi \textsc{Negoro}\altaffilmark{5},
  and
  Ryoji \textsc{Matsumoto}\altaffilmark{6}}
\altaffiltext{1}{Key Laboratory for Research in Galaxies and Cosmology,
  Shanghai Astronomical Observatory, Chinese Academy of Science, 80
  Nandan Road, Shanghai 200030, China}
\email{kawashima-t@shao.ac.cn}
\altaffiltext{2}{National Astronomical Observatory of Japan, Osawa,
Mitaka, Tokyo 181-8588}
\altaffiltext{3}{School of Physical Sciences, Graduate University of
Advanced Study (SOKENDAI), Shonan Village, Hayama, Kanagawa 240-0193}
\altaffiltext{4}{Department of Physics, Tokyo Institute of Technology,
2-12-1 Ookayama, Meguro-ku, Tokyo 181-8588}
\altaffiltext{5}{Department of Physics, Nihon University, 1-8-14
Kanda-Surugadai, Chiyoda-ku, Tokyo 101-8308} 
\altaffiltext{6}{Department of Physics, Graduate School of Science,
Chiba University, 1-33 Yayoi-cho, Inage-ku, Chiba 263-8522}


\maketitle

\begin{abstract}

The tidal disruption event by a supermassive
black hole in Swift J1644+57 can trigger limit-cycle oscillations
between a supercritically accreting X-ray bright state and
a subcritically accreting X-ray dim state. Time evolution of
the debris gas around a black hole with mass $M=10^{6} {\MO}$ is
studied by performing axisymmetric, two-dimensional radiation
hydrodynamic simulations. We assumed the $\alpha$-prescription of
viscosity, in which the viscous stress is proportional to the total
pressure. The mass supply rate from the outer boundary is
assumed to be ${\dot M}_{\rm supply}=100L_{\rm Edd}/c^2$, 
where $L_{\rm Edd}$ is the
Eddington luminosity, and $c$ is the light speed. Since the
mass accretion rate decreases inward by outflows driven by
radiation pressure, the state transition from a supercritically
accreting slim disk state to a subcritically accreting Shakura-Sunyaev
 disk starts from
the inner disk and propagates outward in a timescale of a day.
The sudden drop of the X-ray flux observed in Swift J1644+57
in August 2012 can be explained by this transition. As long as
${\dot M}_{\rm supply}$ exceeds the threshold for the existence of
a radiation pressure dominant disk, accumulation of the accreting
gas in the subcritically accreting region triggers the transition
from a gas pressure dominant Shakura-Sunyaev disk to
a slim disk.
This transition takes place at $t~{\sim}~50/({\alpha}/0.1)~{\rm days}$
after the X-ray darkening. We expect that if $\alpha > 0.01$,
X-ray emission with luminosity 
$\gtrsim 10^{44}$ ${\rm erg}{\cdot}{\rm s}^{-1}$ and jet 
ejection will revive in Swift J1644+57 in 2013--2014.
\end{abstract}

\section{Introduction}
The unusual transient Swift J16449.3+573541 (hereafter Swift J1644+57)
 found in March 2011 has been interpreted as a tidal disruption event,
which enormously increased the accretion rate onto a supermassive black
 hole in an inactive galactic nucleus at redshift $z=0.35$ and launched
 a relativistic jet with Lorentz factor $\Gamma$ $\sim$ 20 
\citep{2011Natur.476..421B,2011Natur.476..425Z}.

The peak isotropic X-ray luminosity of Swift J1644+57 is 
$L_{\rm iso} \sim 10^{48}$ ${\rm erg}{\cdot} {\rm s}^{-1}$.
 Although the radiative flux measured in the jet rest frame reduces by a
factor $\sim \Gamma^{-2}$, the luminosity in the jet rest frame is still
$\sim 10^{46}$ ${\rm erg}{\cdot}{\rm s}^{-1}$. Since the mass $M$ of the
 central black hole of the 
host galaxy of Swift J1644+57 is estimated to be less than 
$2~{\times} ~10^{7}{\MO}$
from the variation time scale of the X-ray intensity and the empirical
law between the mass of the central black hole and the luminosity of the
galactic bulge 
\citep{2011Natur.476..421B}, the jet rest frame luminosity exceeds the
Eddington luminosity 
$L_{\rm Edd}=1.25~{\times}~10^{38}(M/{\MO})$ ${\rm erg}{\cdot}{\rm s}^{-1}$. 
Therefore,
the mass accretion rate $\dot M$ at the jet launching stage should exceed
the Eddington mass accretion rate $\dot M_{\rm Edd} = L_{\rm Edd}/c^2$
 where $c$ is the 
light speed.  
Subsequently, the X-ray luminosity of Swift J1644+57 decreased
as
$L \propto t^{-5/3}$ (
\citealt{2011Sci...333..203B}, but see also
\citealt{2011ApJ...742...32C} and \citealt{2013arXiv1301.1982T}), which can be
explained by  
decrease of the mass supply rate ${\dot M}_{\rm supply}$ from the stellar debris
\citep{1988Natur.333..523R,1989IAUS..136..543P,1989ApJ...346L..13E}. 
Swift J1644+57 
 gives us an opportunity to study the super-Eddington accretion onto a
 supermassive black hole and the transition from a super-Eddington
 accretion flow to a sub-Eddington accretion flow.

 The X-ray luminosity of Swift J1644+57 dramatically dropped in August
2012, when the luminosity decreased from $L_{\rm iso}$ $\sim$ $10^{44}$ 
${\rm erg}{\cdot}{\rm s}^{-1}$ to that
below the detection limit by the {\it Swift} satellite.
The luminosity just before the transition was obtained by using the
observed X-ray flux $F_{0.3-10{\rm keV}} \sim  10^{-12.5}~
{\rm erg}{\cdot}{\rm cm}^{-2}{\cdot}{\rm s}^{-1}$
\citep{2012ATel.4398....1S} and the luminosity distance 
$d_{\rm L} = 1.88~{\rm Gpc}$ \citep{2011Natur.476..421B}.
Observations by the {\it Chandra}
Satellite in November 2012 showed that the X-ray luminosity is about 
$10^{42}$ ${\rm erg}{\cdot}{\rm s}^{-1}$ \citep{2012ATel.4610....1L}.  
\cite{2013ApJ...767..152Z} proposed that the accretion disk transited from a
supercritically accreting slim disk \citep{1988ApJ...332..646A} which
produces X-ray emitting jets to a geometrically-thin, optically thick 
Shakura-Sunyaev disk \citep[][hereafter SSD]{1973A&A....24..337S}, 
in which jet ejection is shut off.
 This model is motivated by the observations of galactic black hole
candidates, in which jets disappear in high/soft states
\citep{2004MNRAS.355.1105F,2011ApJ...739L..19R}. 
More recently, \cite{2013arXiv1301.1982T} predicted that jets and
associated  X-ray emission will be revived in 2016--22 when the
continuous decrease of the accretion rate triggers the state transition
from a high/soft state to a low/hard state, in which the radio emission from
jets is observed in galactic microquasars \citep{2004MNRAS.355.1105F}.

\begin{figure} [!t]
  \begin{center}
    \FigureFile(65mm,65mm){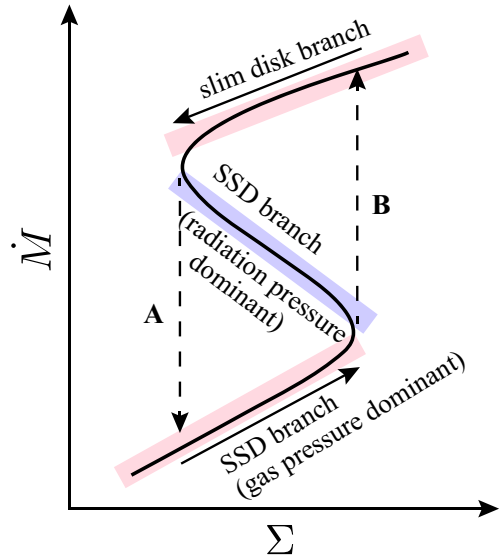}
  \end{center}
  \caption{A Schematic picture of the S-shaped equilibrium curve of 
 black hole accretion disks whose accretion rate is around the Eddington
 critical rate.   
The horizontal axis shows the surface density ${\Sigma}$ 
 of the accretion disk, and the vertical axis represents the accretion rate.
The branch shaded in blue shows the thermally
 and viscously unstable equilibrium curve. 
}\label{fig:s_curve}
\end{figure}

In this Letter, we propose an alternative scenario for the revival of
the jets and X-ray emission in Swift J1644+57 on the basis of the
limit-cycle model of the disk instability which takes place when the
accretion rate is close to the Eddington accretion rate
(e.g., \citealt{1991PASJ...43..147H,2008bhad.book.....K} and references
therein). Figure \ref{fig:s_curve} 
schematically shows the evolutionary track we propose for Swift
J1644+57. When the accretion rate exceeds the Eddington accretion rate,
the accretion flow stays in the upper, slim disk branch
\citep{1988ApJ...332..646A}. As the accretion rate decreases and
becomes less than the critical accretion rate for the existence of the
slim disk branch, the accretion flow will transit to the SSD (transition
denoted by A in Figure \ref{fig:s_curve}). Since the luminosity of the
accretion flow at this transition 
point in Swift J1644+57 is $L$ $\sim$ $1.3{\times}10^{44}$ 
${\rm erg}{\cdot}{\rm s}^{-1}$ and 
$L$ $\sim$ $L_{\rm Edd}$, the black hole mass can be estimated to be
$M=10^{6}{\MO}$, which is consistent with the estimation by \cite{2011Natur.476..421B}. 

The transition A takes place in the thermal timescale of the disk
$t_{\rm th} \sim 100 t_{\rm dyn} \sim 0.01 M/{\MO}$ sec, 
where $t_{\rm dyn}$ is the dynamical
timescale of the innermost region of the disk.
Subsequently, the wave front of the transition propagates in the viscous
timescale 
$t_{\rm vis} \sim t_{\rm th}(H/r)^{-2} \sim (0.01 - 0.1) M/{\MO}$ sec in slim
disks in which $H/r \sim 0.5$, where 
$H$ is the half thickness of the disk.
The luminosity decreases with this timescale.
For the black hole with $10^6 \MO$, the timescale is consistent with
that of the sudden X-ray drop of Swift J1644+57 in August 2012, 
in which the upper limit of the timescale of the X-ray drop constrained
by the ${\it Swift}$ observation is less than $\sim 10^6$ sec.
\cite{2006ApJ...640..923O,2007ApJ...659..205O} showed that the
transition A takes place when ${\dot M}_{\rm supply}$ 
$\lesssim ~100(L_{\rm Edd}/c^2)$.
 Assuming that  ${\dot M}_{\rm supply}$ from the debris of 
the disrupted star decreases as proportional to $t^{-5/3}$
\citep{1989ApJ...346L..13E}, we estimate that ${\dot M}_{\rm supply}$
becomes $\lesssim 100(L_{\rm Edd}/c^2)$ at $\sim$ 1 year after
the maximum mass supply from the debris, which is
 roughly consistent with the duration between the discovery of Swift J1644+57 in
 March 2011 and the drastic decrease of X-ray flux in August 2012. 
In the same way, we estimate that ${\dot M}_{\rm supply}$ 
 exceeds the maximum accretion rate for the existence of
 the gas pressure dominated SSD (${\sim}~L_{\rm Edd}/c^{2}$) for
 ${\sim}$ 10 years.
Therefore, so long as ${\dot M}_{\rm supply}$ exceeds this limit, the
surface density of the gas pressure dominant SSD 
increases, and when the radiation pressure exceeds the gas pressure, 
the disk becomes thermally unstable and transits to
the slim disk (track B).

The accretion rate in the slim disk 
will significantly exceed the Eddington rate because the mass
accumulated in the SSD quickly 
accretes onto the black hole. Radiation hydrodynamic simulations of
supercritical accretion flow onto a black hole 
\citep[e.g.,][]{2005ApJ...628..368O}
showed that radiation pressure driven jets and winds are produced
during the supercritical accretion. Furthermore, radiation hydrodynamic
simulations including Compton cooling 
\citep{2009PASJ...61..769K,2012ApJ...752...18K}
showed that shock heated region formed around the funnel wall of the
radiation pressure supported slim disk (or torus) Comptonizes the soft
photons and emits hard X-rays. Therefore, revival of jets and X-ray
emission is expected when the transition B takes place in Swift
J1644+57.

The limit-cycle behavior in radiation pressure dominant disks was
 demonstrated by two-dimensional radiation hydrodynamic simulations by
 \cite{2006ApJ...640..923O,2007ApJ...659..205O}. 
The limit-cycle can explain the recurrent
bursts in galactic microquasar 
GRS 1915+105 \citep{2003ApJ...596..421W} and IGR J17091-3624
 \citep{2011ApJ...742L..17A}. 
In this Letter, we would like to apply the limit-cycle model 
 to accretion flows onto a supermassive black hole.



\section{Simulation Set-Up}

We solve a set of radiation hydrodynamic equations in spherical
 coordinates 
 ($r$, $\theta$, $\varphi$) assuming axisymmetry. Simulations are
 carried out  by using a global radiation hydrodynamic code
 \citep{2005ApJ...628..368O} with improvements by including the effects
of Compton cooling/heating \citep{2009PASJ...61..769K,2012ApJ...752...18K}.
The radiative transfer is solved by adopting the flux-limited
diffusion approximation \citep{1981ApJ...248..321L,2001ApJS..135...95T}.
The hydrodynamical part is solved by
Virginia Hydrodynamics One (VH-1) code based on the Lagrange-remap
version of piecewise parabolic method \citep{1984JCoPh..54..174C}.
The mass of the black hole is assumed to be  $M=10^6 \MO$.
General relativistic effects are incorporated by a pseudo-Newtonian
potential \citep{1980A&A....88...23P}, 
${\it \Phi}=-GM/(r-r_{\rm s})$ where
$r_{\rm s}$ $( = 2GM/c^2)$ is the Schwarzschild radius, $G$ is the
 gravitational constant. We adopt the $\alpha$-prescription of viscosity
 \citep{1973A&A....24..337S} in which the 
viscous stress is proportional to the total pressure (gas plus
radiation pressure for optically thick plasmas, but gas pressure for
 optically thin limit: see \citealt{2005ApJ...628..368O} in detail) and
 set $\alpha = 0.1$.

The computational domain is 
$2r_{\rm s}$ ${\le}$ $r$ ${\le}$ 
$500r_{\rm s}$ and  
$0$ $\le$ $\theta$ $\le$ ${\pi}/2$.
The number of grid points is ($N_r$, $N_{\theta}$) = (96, 192). 
The grid points are distributed such that $\Delta \ln ~r$ = constant and
$\Delta {\cos}{\theta}=1/N_{\theta}$. 
Mirror symmetry is assumed at the equatorial plane.
The initial state is an isothermal hot ($10^{11}$ K), rarefied,
 optically thin, and spherically symmetric atmosphere. 
Accretion of the stellar
debris is simulated by supplying mass from the outer boundary at
$r=500r_{\rm s}$ near the equatorial plane 
($ 0.45{\pi}$ $\le$ ${\theta}$ $\le$ $0.5{\pi}$)
with a constant rate ${\dot M}_{\rm supply}=100L_{\rm Edd}/c^2$. 
The injected matter is assumed to be cool
(${\sim}10^6$ K) and to have a specific angular momentum 
 corresponding to 
the Keplerian angular momentum at $r=100r_{\rm s}$.
 At the outer boundary except the mass injection region
 (i.e., $r=500r_{\rm s}$ and $0 \le \theta < 0.45{\pi}$), we allow matter
to escape freely but not to enter the computational domain.
 We impose an absorbing boundary condition at $r=2r_{\rm s}$.

\section{Results of Numerical Simulations}

The matter injected from the outer boundary infalls and forms a
torus around $r=100r_{\rm s}$. 
We note that the injected mass accumulates around $60r_{\rm s}$
when the dense disk does not exist in $r < 100r_{\rm s}$ because the ram
pressure of the infalling matter is large enough to push the matter to
the region 60--100$r_{\rm s}$.
As the angular momentum is
distributed by the $\alpha$--viscosity, a geometrically-thin disk is formed.
 As the surface density of the
mass accumulated in the disk exceeds the limit for the existence
for the gas pressure dominant SSD, a supercritically accreting
slim disk is formed. 
The top panel of Figure 2 shows the time evolution of the
isotropic luminosity $L_{\rm iso}$ for the face-on observer
approximately evaluated as
\begin{eqnarray}
L_{\rm iso} = 4{\pi}(r_{\rm out})^{2}F_{r}
(r_{\rm out}, {\theta}_{\rm min}) \label{Liso}, 
\end{eqnarray}
where $r_{\rm out}=500r_{\rm s}$, 
${\theta}_{\rm min} = 0.03({\pi}/2)$,
 and $F_r$ is the radial component of radiative flux 
measured in the observer frame, i.e., 
${\bm F}$ = ${\bm F_{0}}$ + ${\bm v}E_{0}$ + 
${\bm v}{\cdot}{\bf P}_0$. Here, $E_0$, ${\bm F}_0$,
 and ${\bf P}_{0}$ are the radiation energy
  density, the radiation flux,
 and the radiation pressure tensor, which are measured in the 
comoving frame of the fluid, 
respectively.
The bottom panel of Figure 2 shows the mass
accretion rate onto the black hole $\dot M_{\rm acc}$, and the mass outflow
rate ${\dot M}_{\rm out}$ defined as 
\begin{eqnarray}
{\dot M}_{\rm acc} &=& 4{\pi}(r_{\rm in})^{2}{\int}_{0}^{1}
{\rho}(r_{\rm in}, {\theta}) {\max}[-v_r(r_{\rm in}, {\theta}), 0] d{\mu},\\
{\dot M}_{\rm out} &=& 4{\pi}(r_{\rm out})^{2}{\int}_{0}^{1} 
{\rho}(r_{\rm out}, {\theta}) 
{\max}[v_r(r_{\rm out}, {\theta}), 0]d{\mu},  
\end{eqnarray}
where $r_{\rm in}=2r_{\rm s}$,  $\mu = \cos \theta$, and $v_r$ is the radial velocity.

\begin{figure} [!t]
  \begin{center}
    \FigureFile(78mm,78mm){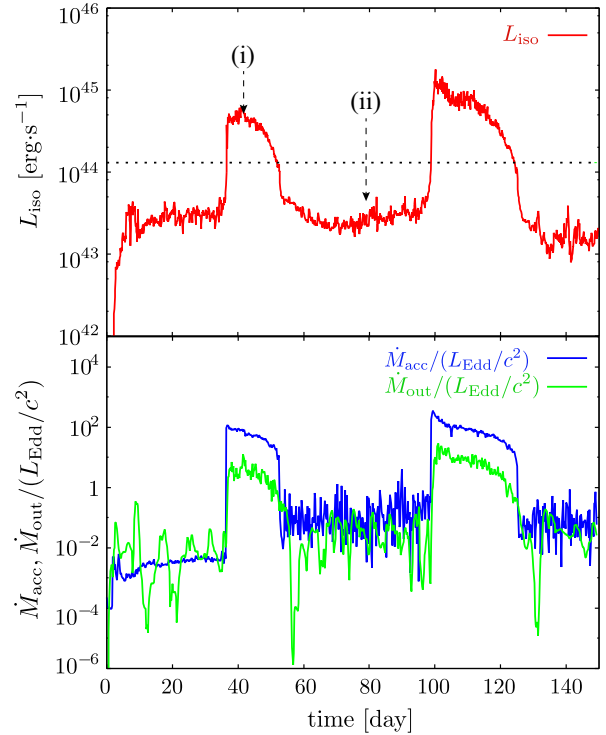}
  \end{center}
  \caption{Time evolution of the isotropic luminosity for the face-on
 observer ({\it top}), the mass accretion rate onto the black hole, and the
 mass outflow rate ({\it bottom}). The dotted line ({\it top}) shows the
 isotropic luminosity of J1644 + 57 when it suddenly 
 dropped in August 2012. (i) indicates the super-Eddington accretion
 stage, and (ii) indicates the sub-Eddington stage.
 }\label{fig:L_curve}
\end{figure}

\begin{figure*} [t]
  \begin{center}
    \FigureFile(170mm,85mm){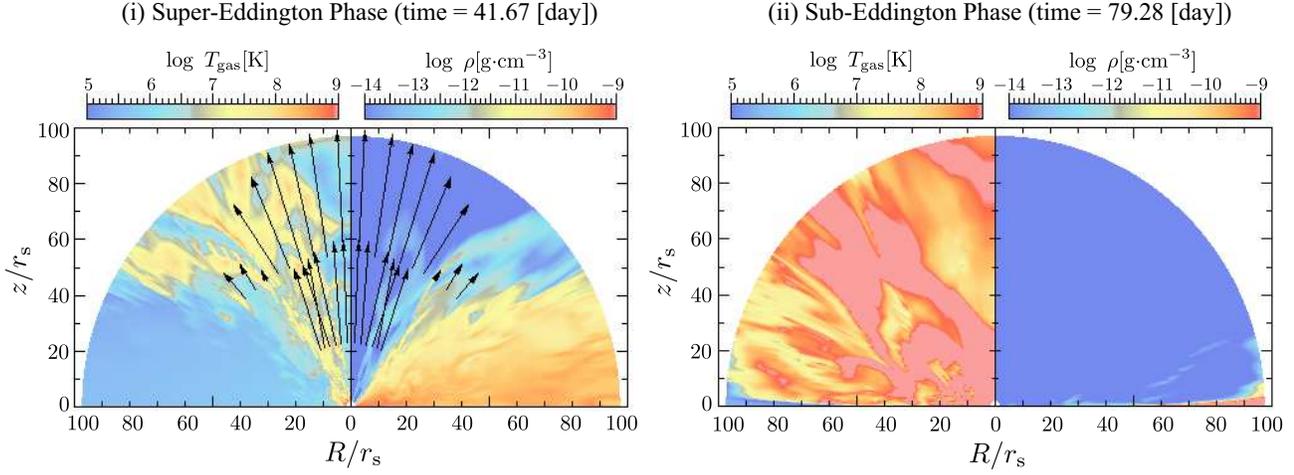}
  \end{center}
  \caption{Snapshots of the simulation in a super-Eddington phase ({\it
 left}) and in a sub-Eddington phase ({\it right}) in $r \le 100r_{\rm
 s}$. The horizontal 
axes
represent $R=r\sin {\theta}$. The arrows
 {\it in the 
 left panel} display the poloidal velocity field of the outflow exceeding
 $10^{-3}c$ at $r~{\simeq}~25r_{\rm s}$ and ${\simeq}~55r_{\rm s}$ in
 logarithmic scale.
The speed for the longest arrow in the figure is $v \sim 0.1c$.}\label{fig:Snap_Shot} 
\end{figure*}

In this simulation, a supercritically accreting disk is formed
 around $t=38$ day. 
Distribution of the temperature, density, and
velocity at this supercritical stage indicated by (i) in
Figure \ref{fig:L_curve} is plotted in the left panel of Figure
 \ref{fig:Snap_Shot}. A radiation 
pressure dominated geometrically thick disk is formed.
The temperature of the disk is $10^{\rm 5-6}$ K, so that UV
radiation will be dominant. 
Above the disk, a radiation pressure
driven jet with speed $\gtrsim$ 0.1$c$ appears around the rotation
 axis and an outflow with $\lesssim$ 0.01$c$ appears outside the jet.
The temperature of the outflow is $10^{7-8.5}$ K. A Hot region
with temperature $\sim$ $10^8$ K also appears around the funnel
wall of the radiation pressure dominant tori near the black
hole \citep[see][]{2012ApJ...752...18K}. 
The UV radiation from the disk is expected to be upscattered to X-rays
 by the inverse Compton scatterings in this hot region. 
The mass accretion rate at this stage
is $10-100L_{\rm Edd}/c^2$, and the isotropic luminosity exceeds
the Eddington luminosity for a black hole with $M=10^6 \MO$.
 The dotted line in Figure \ref{fig:L_curve} shows the luminosity $\sim$
$1.3 \times 10^{44}$ ${\rm erg}{\cdot}{\rm s}^{-1}$, which corresponds
 to the isotropic luminosity of Swift J1644+57 when it suddenly dropped in
 August, 2012. 

 Although the mass supply rate from the outer boundary
is fixed in this simulation, the mass accretion rate decreases
between 40 day and 55 day as plotted in the lower panel of
Figure \ref{fig:L_curve}.
The main reason for the decrease of the mass
accretion rate near the black hole is the mass outflow
in the region 60--300$r_{\rm s}$ during the super-Eddington phase.
We would like to note that the mass outflow rate at $r=500r_{\rm s}$
shown in Figure 2 is smaller than that in 60--300$r_{\rm s}$ because
most of the mass outflowing from this region falls back to
the outer disk.
When the mass accretion rate becomes smaller
than the critical accretion rate for the slim-to-SSD
transition (track A in Figure \ref{fig:s_curve}), the mass accretion
rate and the 
luminosity drastically decreases, and the accretion flow
 transits to a gas pressure dominant, geometrically thin, SSD.
The distribution of the temperature, density, and velocity after
this transition (stage (ii) denoted by a dashed arrow in Figure
\ref{fig:L_curve}) 
 is shown in the right panel in Figure \ref{fig:Snap_Shot}. 
The disk becomes
geometrically-thin, and the jet is shut-off. 
We note that a geometrically thin disk exists down to $R (\equiv r \sin
\theta ) \sim 3 r_{\rm s}$ although
it is not visible in the right panel of Figure \ref{fig:Snap_Shot}.

The mass accretion rate in this SSD is 0.01--0.1$L_{\rm Edd}/c^2$.
Since the mass accretion rate is much smaller than the
mass supply rate $100L_{\rm Edd}/c^2$, the mass accumulates in the
region around $R$ $\sim$ $100r_{\rm s}$. As the surface density of the
disk in this region exceeds the threshold for the  SSD-to-slim transition
(track B in Figure \ref{fig:s_curve}) at 
$t \sim 100$ day in Figure \ref{fig:L_curve}, a supercritically accreting disk
is formed, and radiation pressure driven jets and outflows are revived.


\begin{figure} [!t]
  \begin{center}
    \FigureFile(88mm,88mm){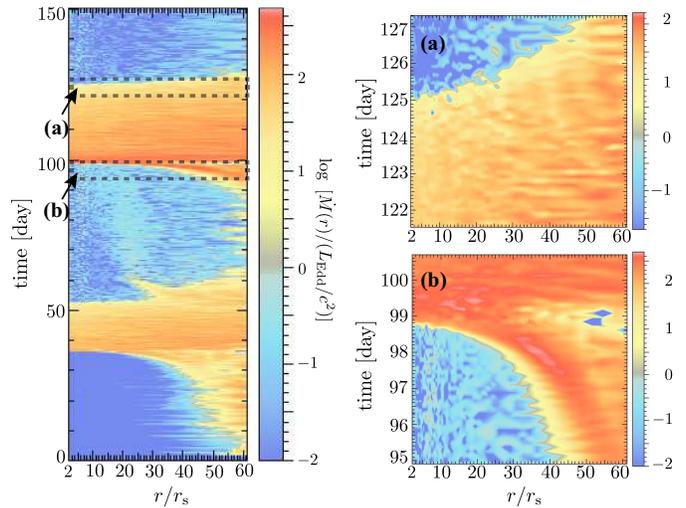}
  \end{center}
  \caption{Time evolution of the ${\dot M}(r)$ and propagation of the
 state transition waves. Color shows ${\dot M}$. {\it Right Panel}
 enlarges the time range denoted by (a) in {\it the left panel}, which
 corresponds to the 
slim-to-SSD transition
and (b) corresponding to the 
SSD-to-slim transition.
}\label{fig:State_Propagation}
\end{figure}

Figure \ref{fig:State_Propagation} shows the propagation of the transition wave between
the slim-to-SSD (orange to blue) and the SSD-to-slim disk
 (blue to orange). Here the mass accretion rate is computed by 
 \begin{eqnarray}
 {\dot M}(r) = 4{\pi}r^{2}{\int}^{1}_{0}{\rho}(r, {\theta})
  {\max}[-v_r(r, {\theta}), 0] d{\mu}. \label{Mdot_r}
 \end{eqnarray}
The slim-to-SSD transition denoted by (a) in Figure
\ref{fig:State_Propagation} starts 
around $t = 125$ day in the inner disk, and the transition completes
in the outer disk at $t = 126$ day. 
The transition starts from the inner region, 
because significant fraction of the accreting mass is ejected by
radiation pressure driven 
outflows, so that accretion rate decreases in the inner region.
Then, the state transition propagates outward in the viscous timescale.
On the other hand, the SSD-to-slim disk transition denoted by (b)
in Figure \ref{fig:State_Propagation} starts in the outer region and
propagates inward. 

\section{Summary and Discussion}
We have shown by two-dimensional axisymmetric radiation hydrodynamic
simulations that when the accretion rate from the debris of a tidally
disrupted star is around
${\dot M}_{\rm supply} \sim 100L_{\rm Edd}/c^2$, recurrent outbursts and jet
ejections take place. 
We assumed the $\alpha$-viscosity in which the viscous stress is
proportional to 
the total pressure. As the mass accretes, supercritically accreting slim
disk and radiation pressure driven jets are formed.  
As the accretion rate near the black hole decreases by this outflow,
the state transition from the slim disk to the geometrically thin
Shakura-Sunyaev disk takes place.
 This transition drastically decreases the luminosity, and turns
 off the jet ejection.

 During the slim disk phase, X-rays can be emitted by the inverse
Compton scatterings of soft photons by hot electrons 
around the funnel wall at the footpoint of the jet and in the outflow.
We computed the Compton {\it y}-parameter for a face-on
 observer by integrating the Thomson opacity from $z = 400 r_{\rm s}$ down
 to the point where the effective optical depth ${\tau}_{\rm eff} = 1$
 for each radius and found that {\it y} $\gg$ 1 during the slim disk
 phase.
The spectral bump around 1keV in Swift J1644+57 observed by {\it Swift}
\citep{Saxton2012} may be explained by the Comptonization of UV disk
photons in the Compton-thick outflow.  
On the other hand, during the SSD phase, it is found that the hot region
in the disk corona mostly does not contribute to the luminosity except
small regions around $R = 5r_{\rm s}$, where {\it y} sometimes
exceeds unity. 
Therefore, we expect that the radiation at this stage is mostly emitted
in UV, and the X-ray luminosity is small.
The inverse Compton scatterings in the small region 
where {\it y} $> 1$ may
be the origin of the X-ray emission with luminosity $L_{\rm X} \sim 10^{42} {\rm erg}{\cdot}{\rm s}^{-1}$ 
observed by the {\it Chandra} satellite in November, 2012. 

The duration of 
the sub-Eddington accretion is $\sim 50$ days
when $\alpha =0.1$. We employed $\alpha = 0.1$ to save the computational time.
Three-dimensional MHD simulations of the growth of the magneto-rotational
instability in radiatively inefficient disk indicate that 
$0.01 < \alpha < 0.1$
\citep[e.g.,][]{2000ApJ...528..462H,2004PASJ...56..671M,2011ApJ...738...84H}
.
Three-dimensional local radiation MHD simulations indicates that
$\alpha > 0.01$ \citep{2009ApJ...704..781H,2009ApJ...691...16H}. 
If we assume $\alpha = 0.01$, 
the duration of the subcritical accretion phase is $\sim 500$ days
because the viscous timescale is proportional to $\alpha^{-1}$.
We expect, therefore that the revival of X-ray emission and jet
ejection will take place within $\sim 500$ days after the sudden
darkening in August 2012.

Finally, let us discuss the difference of observational features between the
jet revival models for Swift J1644+57, i.e., between the
model proposed by \cite{2013arXiv1301.1982T} 
and our alternative model.
When the jet is re-launched, the super-Eddington accretion
will take place again in our model, while the radiatively inefficient
accretion flow will appear in the model by \cite{2013arXiv1301.1982T}. 
According to our model, the photon spectral shape will be similar to
that before the dramatic darkening, and the
luminosity will exceed the Eddington luminosity again. 
On the other hand, according to the model by
\cite{2013arXiv1301.1982T}, the spectral state is expected to be similar to that
of lower-luminosity blazars, 
and the luminosity
measured in the jet rest frame
will be significantly lower than the
Eddington luminosity. 
In addition, the time at which the revival of the jet starts is earlier
in our model than in their model.
In subsequent papers, we would like to confirm the timescale for the
revival of the jet by carrying out simulations without assuming 
${\dot M}_{\rm supply} = $ constant.
The spectral calculations by post-processing the simulation data
also remain  as a future work.

\bigskip

The numerical simulations were
carried out on the XT4 at the Center for Computational Astrophysics,
National Astronomical Observatory of Japan. This
work was supported by the JSPS Grants in Aid for Scientific Research
(24740127, K.O; 23340040, R.M.).


\begin{thebibliography}{34}
\expandafter\ifx\csname natexlab\endcsname\relax\def\natexlab#1{#1}\fi

\bibitem[{{Abramowicz} {et~al.}(1988){Abramowicz}, {Czerny}, {Lasota}, \& {Szuszkiewicz}}]{1988ApJ...332..646A}
{Abramowicz}, M.~A., {Czerny}, B., {Lasota}, J.~P., \& {Szuszkiewicz}, E. 1988,
  \apj, 332, 646

\bibitem[{{Altamirano} {et~al.}(2011){Altamirano}, {Belloni}, {Linares}, {van
  der Klis}, {Wijnands}, {Curran}, {Kalamkar}, {Stiele}, {Motta},
  {Mu{\~n}oz-Darias}, {Casella}, \& {Krimm}}]{2011ApJ...742L..17A}
{Altamirano}, D., {Belloni}, T., {Linares}, M., {et~al.} 2011, \apjl, 742, L17

\bibitem[{{Bloom} {et~al.}(2011){Bloom}, {Giannios}, {Metzger}, {Cenko},
  {Perley}, {Butler}, {Tanvir}, {Levan}, {O'Brien}, {Strubbe}, {De Colle},
  {Ramirez-Ruiz}, {Lee}, {Nayakshin}, {Quataert}, {King}, {Cucchiara},
  {Guillochon}, {Bower}, {Fruchter}, {Morgan}, \& {van der
  Horst}}]{2011Sci...333..203B}
{Bloom}, J.~S., {Giannios}, D., {Metzger}, B.~D., {et~al.} 2011, Science, 333,
  203

\bibitem[{{Burrows} {et~al.}(2011){Burrows}, {Kennea}, {Ghisellini}, {Mangano},
  {Zhang}, {Page}, {Eracleous}, {Romano}, {Sakamoto}, {Falcone}, {Osborne},
  {Campana}, {Beardmore}, {Breeveld}, {Chester}, {Corbet}, {Covino},
  {Cummings}, {D'Avanzo}, {D'Elia}, {Esposito}, {Evans}, {Fugazza}, {Gelbord},
  {Hiroi}, {Holland}, {Huang}, {Im}, {Israel}, {Jeon}, {Jeon}, {Jun}, {Kawai},
  {Kim}, {Krimm}, {Marshall}, {P.~M{\'e}sz{\'a}ros}, {Negoro}, {Omodei},
  {Park}, {Perkins}, {Sugizaki}, {Sung}, {Tagliaferri}, {Troja}, {Ueda},
  {Urata}, {Usui}, {Antonelli}, {Barthelmy}, {Cusumano}, {Giommi}, {Melandri},
  {Perri}, {Racusin}, {Sbarufatti}, {Siegel}, \&
  {Gehrels}}]{2011Natur.476..421B}
{Burrows}, D.~N., {Kennea}, J.~A., {Ghisellini}, G., {et~al.} 2011, \nat, 476,
  421

\bibitem[{{Cannizzo} {et~al.}(2011){Cannizzo}, {Troja}, \&
  {Lodato}}]{2011ApJ...742...32C}
{Cannizzo}, J.~K., {Troja}, E., \& {Lodato}, G. 2011, \apj, 742, 32

\bibitem[{{Colella} \& {Woodward}(1984)}]{1984JCoPh..54..174C}
{Colella}, P., \& {Woodward}, P.~R. 1984, Journal of Computational Physics, 54,
  174

\bibitem[{{Evans} \& {Kochanek}(1989)}]{1989ApJ...346L..13E}
{Evans}, C.~R., \& {Kochanek}, C.~S. 1989, \apjl, 346, L13

\bibitem[{{Fender} {et~al.}(2004){Fender}, {Belloni}, \&
  {Gallo}}]{2004MNRAS.355.1105F}
{Fender}, R.~P., {Belloni}, T.~M., \& {Gallo}, E. 2004, \mnras, 355, 1105

\bibitem[{{Hawley}(2000)}]{2000ApJ...528..462H}
{Hawley}, J.~F. 2000, \apj, 528, 462

\bibitem[{{Hawley} {et~al.}(2011){Hawley}, {Guan}, \&
  {Krolik}}]{2011ApJ...738...84H}
{Hawley}, J.~F., {Guan}, X., \& {Krolik}, J.~H. 2011, \apj, 738, 84

\bibitem[{{Hirose} {et~al.}(2009{\natexlab{a}}){Hirose}, {Blaes}, \&
  {Krolik}}]{2009ApJ...704..781H}
{Hirose}, S., {Blaes}, O., \& {Krolik}, J.~H. 2009{\natexlab{a}}, \apj, 704,
  781

\bibitem[{{Hirose} {et~al.}(2009{\natexlab{b}}){Hirose}, {Krolik}, \&
  {Blaes}}]{2009ApJ...691...16H}
{Hirose}, S., {Krolik}, J.~H., \& {Blaes}, O. 2009{\natexlab{b}}, \apj, 691, 16

\bibitem[{{Honma} {et~al.}(1991){Honma}, {Kato}, \&
  {Matsumoto}}]{1991PASJ...43..147H}
{Honma}, F., {Kato}, S., \& {Matsumoto}, R. 1991, \pasj, 43, 147

\bibitem[{{Kato} {et~al.}(2008){Kato}, {Fukue}, \&
  {Mineshige}}]{2008bhad.book.....K}
{Kato}, S., {Fukue}, J., \& {Mineshige}, S. 2008, {Black-Hole Accretion Disks
  --- Towards a New Paradigm --- (Kyoto: Kyoto University Press)}

\bibitem[{{Kawashima} {et~al.}(2009){Kawashima}, {Ohsuga}, {Mineshige},
  {Heinzeller}, {Takabe}, \& {Matsumoto}}]{2009PASJ...61..769K}
{Kawashima}, T., {Ohsuga}, K., {Mineshige}, S., {et~al.} 2009, \pasj, 61, 769

\bibitem[{{Kawashima} {et~al.}(2012){Kawashima}, {Ohsuga}, {Mineshige},
  {Yoshida}, {Heinzeller}, \& {Matsumoto}}]{2012ApJ...752...18K}
---. 2012, \apj, 752, 18

\bibitem[{{Levan} \& {Tanvir}(2012)}]{2012ATel.4610....1L}
{Levan}, A.~J., \& {Tanvir}, N.~R. 2012, The Astronomer's Telegram, 4610, 1

\bibitem[{{Levermore} \& {Pomraning}(1981)}]{1981ApJ...248..321L}
{Levermore}, C.~D., \& {Pomraning}, G.~C. 1981, \apj, 248, 321

\bibitem[{{Machida} {et~al.}(2004){Machida}, {Nakamura}, \&
  {Matsumoto}}]{2004PASJ...56..671M}
{Machida}, M., {Nakamura}, K., \& {Matsumoto}, R. 2004, \pasj, 56, 671

\bibitem[{{Ohsuga}(2006)}]{2006ApJ...640..923O}
{Ohsuga}, K. 2006, \apj, 640, 923

\bibitem[{{Ohsuga}(2007)}]{2007ApJ...659..205O}
---. 2007, \apj, 659, 205

\bibitem[{{Ohsuga} {et~al.}(2005){Ohsuga}, {Mori}, {Nakamoto}, \&
  {Mineshige}}]{2005ApJ...628..368O}
{Ohsuga}, K., {Mori}, M., {Nakamoto}, T., \& {Mineshige}, S. 2005, \apj, 628,
  368

\bibitem[{{Paczy{\'n}sky} \& {Wiita}(1980)}]{1980A&A....88...23P}
{Paczy{\'n}sky}, B., \& {Wiita}, P.~J. 1980, \aap, 88, 23

\bibitem[{{Phinney}(1989)}]{1989IAUS..136..543P}
{Phinney}, E.~S. 1989, in IAU Symposium, Vol. 136, The Center of the Galaxy,
  ed. M.~{Morris}, 543

\bibitem[{{Rees}(1988)}]{1988Natur.333..523R}
{Rees}, M.~J. 1988, \nat, 333, 523

\bibitem[{{Russell} {et~al.}(2011){Russell}, {Miller-Jones}, {Maccarone},
  {Yang}, {Fender}, \& {Lewis}}]{2011ApJ...739L..19R}
{Russell}, D.~M., {Miller-Jones}, J.~C.~A., {Maccarone}, T.~J., {et~al.} 2011,
  \apjl, 739, L19

\bibitem[{Saxton {et~al.}(2012)Saxton, Soria, Wu, \& Kuin}]{Saxton2012}
Saxton, C.~J., Soria, R., Wu, K., \& Kuin, N. P.~M. 2012, Monthly Notices of
  the Royal Astronomical Society, 422, 1625

\bibitem[{{Sbarufatti} {et~al.}(2012){Sbarufatti}, {Burrows}, {Gehrels}, \&
  {Kennea}}]{2012ATel.4398....1S}
{Sbarufatti}, B., {Burrows}, D.~N., {Gehrels}, N., \& {Kennea}, J.~A. 2012, The
  Astronomer's Telegram, 4398, 1

\bibitem[{{Shakura} \& {Sunyaev}(1973)}]{1973A&A....24..337S}
{Shakura}, N.~I., \& {Sunyaev}, R.~A. 1973, \aap, 24, 337

\bibitem[{{Tchekhovskoy} {et~al.}(2013){Tchekhovskoy}, {Metzger}, {Giannios},
  \& {Kelley}}]{2013arXiv1301.1982T}
{Tchekhovskoy}, A., {Metzger}, B.~D., {Giannios}, D., \& {Kelley}, L.~Z. 2013,
  arXiv:1301.1982

\bibitem[{{Turner} \& {Stone}(2001)}]{2001ApJS..135...95T}
{Turner}, N.~J., \& {Stone}, J.~M. 2001, \apjs, 135, 95

\bibitem[{{Watarai} \& {Mineshige}(2003)}]{2003ApJ...596..421W}
{Watarai}, K., \& {Mineshige}, S. 2003, \apj, 596, 421

\bibitem[{{Zauderer} {et~al.}(2013){Zauderer}, {Berger}, {Margutti}, {Pooley},
  {Sari}, {Soderberg}, {Brunthaler}, \& {Bietenholz}}]{2013ApJ...767..152Z}
{Zauderer}, B.~A., {Berger}, E., {Margutti}, R., {et~al.} 2013, \apj, 767, 152

\bibitem[{{Zauderer} {et~al.}(2011){Zauderer}, {Berger}, {Soderberg}, {Loeb},
  {Narayan}, {Frail}, {Petitpas}, {Brunthaler}, {Chornock}, {Carpenter},
  {Pooley}, {Mooley}, {Kulkarni}, {Margutti}, {Fox}, {Nakar}, {Patel},
  {Volgenau}, {Culverhouse}, {Bietenholz}, {Rupen}, {Max-Moerbeck}, {Readhead},
  {Richards}, {Shepherd}, {Storm}, \& {Hull}}]{2011Natur.476..425Z}
{Zauderer}, B.~A., {Berger}, E., {Soderberg}, A.~M., {et~al.} 2011, \nat, 476,
  425

\end{thebibliography}



\end{document}